\documentclass[twocolumn,english,aps,prx,floatfix,amssymb,superscriptaddress,longbibliography]{revtex4}
\usepackage[latin9]{inputenc}
\usepackage{verbatim}
\usepackage{float}
\usepackage{amsmath}
\usepackage{amssymb}
\usepackage{graphicx}
\usepackage{color}
\usepackage{xcolor}
\usepackage{tikz}
\newcommand{\ket}[1]{\ensuremath{\left| #1 \right>}}

\usepackage{bbold}

\usepackage[bookmarks=false,linkcolor=blue,urlcolor=blue,colorlinks,citecolor=blue]{hyperref}
\makeatletter

\newcommand{\be}{\begin{equation}}
\newcommand{\ee}{\end{equation}}
\newcommand{\bea}{\begin{eqnarray}}
\newcommand{\eea}{\end{eqnarray}}

\begin{document}

\title{Beyond Integrability Preserving Renormalization-Group Protocol in Non-Hermitian Hamiltonians with Time-Dependent Interaction Strengths}
\author{Parameshwar R. Pasnoori}
\affiliation{Condensed Matter Theory Center, Department of Physics, University of Maryland, College Park, MD 20742, USA}
\affiliation{Department of Physics and Astronomy, University of California, Los Angeles, CA 90095, USA}

\email{parmesh12@g.ucla.edu}
\begin{abstract}
It is well established that in time-dependent quantum systems, integrability preserving time-dependent interaction strengths are identical to the renormalization group (RG) trajectories of the corresponding static model when time `$t$' in the driven model is identified with the logarithm of the cutoff `$\log\Lambda$' of the static model. We refer to this integrability preserving driving as the RG protocol. In this work we extend the class of time-dependent integrable models to include non-Hermitian quantum models with time-dependent interaction strengths. Using the recently formulated generalized Bethe ansatz framework [P. R. Pasnoori, Phys. Rev. B 112, L060409 (2025)],  we show that the constraints imposed by integrability are more general: The interaction strengths of the static model that flow in the RG follow the respective RG trajectories in the corresponding time-dependent model as described above. In addition, the interaction strengths of the static model that are RG invariant can either be constant or have a specific time-dependence in the corresponding time-dependent model which is constrained by integrability. Thus we establish that in the context of time-dependent non-Hermitian systems, the set of integrability preserving time-dependent strengths is larger than the set corresponding to the RG protocol.

\end{abstract}
\maketitle
\paragraph{Introduction}
 Integrability provides a rigorous mathematical framework to study strongly interacting many-body systems. Bethe ansatz \cite{Bethe1931,Hulthen,SklyaninQISM} which is the framework that provides exact solutions of integrable field theories and statistical models \cite{AndreiLowenstein79,Thacker,Japaridze,JimboMiwa1995,sixV,eightV,LiebFu,ZAMOLODCHIKOV,ZAMOLODCHIKOVAA,Sarkar,FKor,IZERGIN,LL1,LL2,McGuire,GAUDIN,CNYang,Sutherland,RICHARDSON,Gaudin1976,Dorey,BanksZaks1982,Novikov1983,Polchinski1988}, and provides tools to study novel phenomena in both closed and open quantum systems ranging from symmetry breaking phases \cite{TerrasXXZ1,TerrasXXZ2,XXZpaper,XXZPD}, symmetry protected topological phases \cite{PAA1,PAA2,pasnoori2025duality,pasnoori2025interplay,circuitSPT}, non perturbative effects of quantum impurities \cite{Andrei80,AndreiLowensein81,Wiegmann_1981,parmeshkondo1,parmeshkondo2,TsvelickWiegmann1983,KondoXXX} and dissipative effects in strongly interacting systems \cite{NHK1,NHK2,PTkondo,PTXXX,FabianNHH}. Recently, there has been a surge in the study of systems with time-dependent strengths \cite{Benhoare,PasnooriRG,komatsu2026,gamayun,Chernyak_2021,sinitsyn2026,Barik2026,timeBCS,TCSun,multiLZ,Sinitsyn,YUZBASHYAN,PasnooriKondo,PasnooriGrossNeveu,pasnoorikondo2,pasnoorigrossneveu2} partly owing to the advancements in the experimental control over the parameters of the Hamiltonian \cite{RAIZENcold,coldatomsBA,SPTcold,Zhangcold} and also due to wide spread applications in quantum information science \cite{timeSCcircuit,timetrotter,timesimulationclock,circuitHamiltonian,Cao_2025}. To obtain exact solutions of Hamiltonians with time-dependent interaction strengths, a generalized Bethe ansatz framework was developed in \cite{PasnooriKondo}. This framework provides constraints on the time-dependent interaction strengths for the system to be integrable. For interaction strengths satisfying these constraints, the framework provides exact many-body wavefunction that satisfies the time-dependent Schrodinger equation \cite{PasnooriKondo,PasnooriGrossNeveu,pasnoorikondo2,pasnoorigrossneveu2}. Prior to the development of the generalized Bethe ansatz framework, in a ground breaking work B. Hoare, N. Levine and A. A. Tseytlin \cite{Benhoare} have shown that integrability is preserved with time dependent couplings provided the time-dependent theory admits a Lax connection where the time-dependent coupling strengths satisfy the one-loop RG equations of the time-independent theory. Using the generalized Bethe ansatz framework, it was later shown in \cite{PasnooriRG} that integrable models with constant interaction strengths that are solvable by the Bethe ansatz method remain integrable when the interaction strengths are time-dependent provided they satisfy certain constraint conditions. For small values of the coupling strengths, these conditions were shown to coincide with the RG equations of the corresponding static model thus corroborating the results found in \cite{Benhoare}. We refer to the driving where the time-dependent coupling strengths follow the RG trajectories thereby preserving integrability as the RG protocol. It was recently shown in \cite{komatsu2026} that one can construct time-dependent integrable field theories by generalizing the four dimensional Chern-Simons theory extending the seminal work relating gauge theories and integrability by K. Costello, E. Witten and M. Yamazaki \cite{costello1,costello2,costello3}, where they have found the same connection between the time-dependent coupling strengths and the RG flow mentioned above.

In this work we question whether the connection between the time-dependent couplings and the RG flow fundamental or are the constraints imposed by integrability fundamental. This question is certainly deep, but one can find an answer by asking a simpler question: can the set of integrability preserving time-dependent strengths be larger than the set of time-dependent strengths corresponding to the RG protocol. We find an affirmative answer to this question when the time-dependent coupling strengths are allowed to be complex. As a concrete example, we consider the most simplistic strongly interacting Hamiltonian with complex time-dependent interaction strengths, namely the time dependent $SU(2)$ non-Hermitian Kondo model, and solve it using the generalized Bethe ansatz method \cite{PasnooriKondo}. We find that the integrability imposes constraints such that the coupling strengths corresponding to the static model that flow in the RG necessarily have to follow the same trajectories in the time dependent model in agreement with the aforementioned results. In addition, we find that integrability imposes constraints on the time-dependent coupling strengths whose counterparts in the corresponding static model are RG invariant. We show that these constraints cannot only be satisfied by constant strengths which simply corresponds to the RG protocol, but in addition, the constraints allow these time-dependent strengths to take specific functional forms. Thereby demonstrating that the set of integrability preserving time-dependent coupling strengths is larger than the set corresponding to the RG protocol.

\paragraph{The model}
The Hamiltonian of the time-dependent non Hermitian Kondo model is given by
(\ref{Hamiltonian})
\bea \nonumber H=\int_{-L/2}^{L/2} \; dx\;  \big\{\Psi^{\dagger}_{a}(x)(-i\partial_x)\Psi_{a}(x) \\+ J(t) \Psi^{\dagger}_{a}(0)\left(\vec{\sigma}_{ab}\cdot\vec{S}_{\alpha\beta}\right)\Psi_{b}(0)\big\}, \label{Hamiltonian}
\eea
where $\Psi_{a}(x)$ describes the fermion (electron) field with subscript $a=\uparrow,\downarrow$ denoting the spin. $S$ represents the impurity and $J(t)\in \mathbb C$ is the time dependent interaction strength. For simplicity, we have set the Fermi velocity $v_F=1$. This model with constant interaction strengths has been shown to be experimentally realizable \cite{NHK1} in a dissipative AMO system consisting of two-orbital $^{173}$Yb gas atoms, where the atoms in the metastable excited state play the role of spin S = 1/2 impurities and the atoms in the ground state play the role of itinerant electrons. Since the purpose of the current work is to understand the integrability structure of (\ref{Hamiltonian}), we ignore the details of the experimental realization and study the model using the generalized Bethe ansatz framework. Before we venture into the solution of the time-dependent model and discuss how the set of integrability preserving coupling strengths encompass the set corresponding to the RG protocol, it is important to understand the RG invariants and the phase structure of the static non Hermitian Kondo model. We shall briefly discuss this below.

\paragraph{RG invariants in the static model}
  
The model (\ref{Hamiltonian}) with constant interaction strengths has been solved using the Bethe ansatz \cite{NHK1}, where it was shown to exhibit different phases characterized by the nature of the impurity screening or lack thereof. Analyzing the RG flow it was argued that a phase transition occurs at a certain critical value of the interaction strength, where on either side of this phase transition, the system flows to distinct fixed points which correspond to a Kondo phase and a non Kondo phase. In a later work \cite{NHK2}, a more thorough Bethe ansatz analysis was performed where it was shown that in addition to the two phases mentioned above, there exists an intermediate phase named $\widetilde{YSR}$. We very briefly discuss the results of this paper in the following.

\begin{center}
\begin{figure}
\includegraphics[width=0.9\columnwidth]{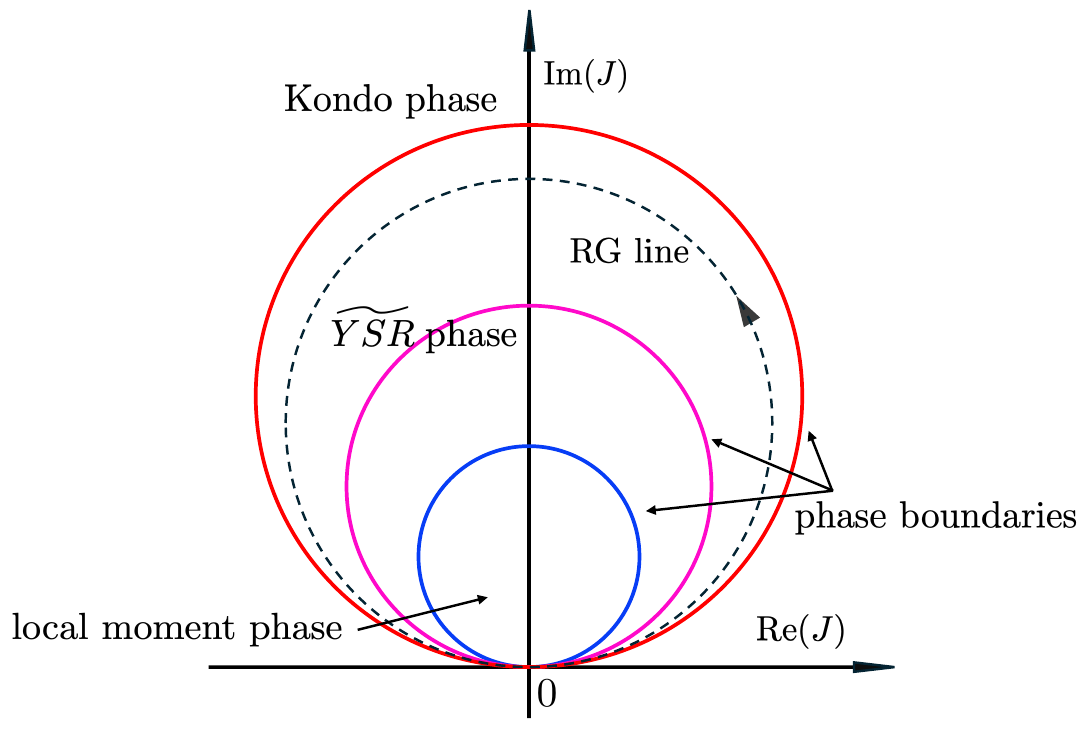}
\caption{Figure depicts the three phases exhibited by the static non-Hermitian Kondo model: the Kondo phase, the $\widetilde{YSR}$ phase and the local moment phase. The x and y-axes correspond to the real and the imaginary values of the interaction strength respectively. The trajectories of the RG flow are perfect circles resting at the origin as shown (dashed circles) with the arrow representing the flow in the UV. All the RG trajectory circles are centered on the imaginary axis where the radius of the circle corresponds to the respective RG invariant. The phase boundaries occur at specific values of the RG invariant values and are represented by solid circles.  }
\label{fig:picture1}
\end{figure}
\end{center}

There exists an important parameter `$f$' which is related to the bare interaction strength $J$ through the relation $f=\frac{1}{2J}\left(1-\frac{3J^2}{4}\right)$. Note that this relation is non-universal. A different regularization scheme results in a different expression. Both of them coincide in the universal regime which corresponds to small values of $J$, where $f\sim 1/J$. The negative of the imaginary part of $f$ which we label $f_i$ is an RG invariant and measures the departure from Hermiticity whereas its real part which we label $f_r$ flows along RG. The system exhibits three distinct phases depending on the values of $f_i$: The Kondo phase exists for $f_i<1/2$ where the impurity is screened by the many-body Kondo effect which is characterized by the generation of the strong coupling scale, the Kondo temperature $T_K$ given by
\begin{align}T_K=2\Lambda e^{-\pi f_r}.\label{kondotemp}
\end{align}
One takes the scaling limit where the physical scale $T_K$ is held fixed while the cutoff $\Lambda\rightarrow \infty$. This requires one to take the parameter $f_r\rightarrow \infty$.

As the strength of non-Hermiticity is increased such that for $1/2<f_i<3/2$, the system exhibits $\widetilde{YSR}$ phase where the system ceases to generate a many-body Kondo effect and instead exhibits a single particle bound state which screens the impurity similar to the Yu-Shiba-Rusinov $YSR$ states that arise when a magnetic impurity is coupled to a superconductor \cite{Yu, Shiba,Rusinov,parmeshkondo2}. But unlike the YSR states, here the bulk of the system is gapless but yet, the system exhibits a `stable' bound state whose real part of the energy is given by
\begin{align} E_B=-T_K \sin \pi f_i.
\end{align}
Thus the state in which the impurity is screened has lower real part of the energy for $1/2<f_i<1$ which we refer to as the ground state. The bound state can be removed, and thus one can unscreen the impurity which costs energy and corresponds to an excited state. For $1<f_i<3/2$, the bound state's real part of the energy is positive and hence the state in which the bound state is present, that is the state in which the impurity is screened is an excited state. Whereas, the state in which the bound state is absent, that is the state in which the impurity is unscreened is the ground state. Hence, at $f_i=1$ the system exhibits a first order phase transition where the ground state of the system changes. The imaginary part of the bound state's energy characterizes its life time and is given by $E_i=T_K\cos \pi f_i <0$. Thus implying that the bound state has a finite lifetime. For larger departures from Hermiticity where the values of the parameter $f_i>3/2$, the system enters local moment phase, where the bound state which is responsible for the screening of the impurity no longer exists and the impurity remains unscreened in the ground state \footnote{Similar to the case of the Kondo impurity coupled to a one dimensional superconductor \cite{parmeshkondo1,parmeshkondo2}, one may expect that the impurity is actually partially screened due to the remnant anti-ferromagnetic interactions. As the values of $f_i$ are increased, the impurity screening is reduced until it becomes completely free for much larger values of $f_i$.}. 

The RG invariant and the dynamically generated scale $T_K$ described above can be inferred from the one loop order RG equation, which is given by

\begin{align} \frac{d}{d\log\Lambda}\left(\frac{1}{J}\right)=\frac{1}{\pi}. \label{rg}\end{align}

One can observe that only the real part of the inverse of the interaction strength $1/J$ flows, whereas its imaginary part is a constant. Thus one can infer that $Im(1/J)=-f_i$ is an RG invariant, whereas $Re(1/J)=f_r$ flows under RG which gives rise to the strong coupling scale discussed above. Typically one plots $Im(J)$ as a function of the $Re(J)$, in which case one obtains perfect circles called limit cycles that are centered on the imaginary axis as shown in the figure (\ref{fig:picture1}). The flow in the UV (increasing the cutoff $\Lambda$) is along the counter-clockwise direction. It is interesting to note that the RG invariant is related to the radius $R$ of the circle, or equivalently, the center of the circle where $2R=|1/Im(1/J)|$.

\paragraph{Generalized Bethe ansatz solution of the time-dependent model}
Now let us consider the time-dependent model (\ref{Hamiltonian}). This model can be solved using the generalized Bethe ansatz framework developed in \cite{PasnooriKondo}. In this framework one constructs an exact wavefunction, which is the solution to the time-dependent Schrodinger equation. The wavefunction consists of amplitudes which correspond to different ordering of the particles with respect to each other and also with respect to the impurity. The amplitudes are related to one another through various S-matrices, which unlike the regular Bethe ansatz solvable systems, depend on time and also on the spatial coordinates of the particles. Applying periodic boundary conditions, one obtains a set of matrix difference equations acting in the spin spaces of the particles and the impurity, which are called the quantum Knizhnik-Zamolodchikov (qKZ) equations \cite{Smirnov_1986,Frenkel,rishetikhin1,Babujian_1997,rishetikhin2,varchenko,Tarasov:1994bb}. The consistency of the solution imposes constraints on the time-dependent interaction strengths. For interaction strengths satisfying these constraints, the system is integrable, and the solution to the qKZ equations provides the explicit form of the many-body wavefunction, using which one can extract the exact dynamics \cite{pasnoorigrossneveu2}. Here we skip the details of the construction of the wavefunction and only provide an overview to be self consistent. 

The number of particles $N =\sum_{a}\int_{-L/2}^{L/2} dx\:  \Psi^{\dagger}_{a}(x)\Psi_{a}(x)$ is a conserved quantity, hence we can look for wave function labeled by $N$ which satisfies the time dependent Schrodinger equation

\bea i\partial_t \ket{\Psi_N}= H \ket{\Psi_N},\label{SEk}\eea

where $H$ is the Hamiltonian (\ref{Hamiltonian}). The wavefunction takes the form
\begin{align}\label{npform}\ket{\Psi_N}\hspace{-0.04in}=\hspace{-0.15in}\sum_{\sigma_1...\sigma_N}\hspace{-0.7mm}\prod_{j=1}^{N}\int_{-L/2}^{L/2}\hspace{-0.1in}dx_j \Psi^{\dagger}_{\sigma_j}(x_j)\mathcal{A}F_{\sigma_1...\sigma_N\alpha}(x_1,...,x_N,t)\ket{\alpha}, 
\end{align}
where $\sigma_j$ denote the spin indices of the electrons and $\alpha$ denotes the spin of the impurity and $\mathcal{A}$ denotes anti-symmetrization with respect to $x_i$ and $\sigma_i$. Using the above expression (\ref{npform}) in the Schrodinger equation (\ref{SEk}), one obtains the following equation

\begin{align} &\nonumber-i(\partial_t+\sum_{j=1}^N\partial_{x_j})\mathcal{A}F_{\sigma_1...\sigma_N\alpha}(x_1,...,x_N,t)
\\\nonumber&+J(t)\big(\sum_{j=1}^N\delta(x_j)\:\vec{\sigma}_{\sigma_j\gamma_j}\cdot\vec{S}_{\alpha\beta}\:\prod_{l=1,l\neq j}^NI_{\sigma_l\gamma_l}\big)\\&\;\;\;\;\;\;\;\;\;\times \mathcal{A}F_{\gamma_1...\gamma_N\beta}(x_1,...,x_N,t)=0.\label{npdiffk}\end{align}
As mentioned above, one has to distinguish between the amplitudes corresponding to the particle being on the left or right sides of the impurity and also between the amplitudes corresponding to different ordering of particles with respect to each other as well. We have 
\begin{align} 
F_{\sigma_1...\sigma_N}(x_1,...,x_N,t)
= \sum_Q \theta(\{x_{Q(j)}\})  f^Q_{\sigma_1...\sigma_N}(z_1,...,z_N),\label{npformexp}
\end{align}
where we used the notation $z_i=x_i-t$, $i=1,..,N$. In this expression, $Q$ denotes a permutation of the position orderings of particles and  $\theta(\{x_{Q(j)}\})$ is the Heaviside function that vanishes unless $x_{Q(1)} \le \dots \le x_{Q(N)}$. Here $f^{Q}_{\sigma_1...\sigma_N} (z_1,...,z_N)$ is the amplitude corresponding to the ordering of the particles denoted by $Q$.  Applying periodic boundary conditions in the spatial direction results in the following relations

\be f^{j...0}_{\sigma_1...\sigma_N}(z_1,...,z_j,...,z_N)=f^{...0j}_{\sigma_1...\sigma_N}(z_1,...,z_j+L,...,z_N).\label{pbcnp}\ee
 Here ``..." in the superscripts corresponds to any ordering of the rest of the particles, which is the same in the amplitudes on both sides of the equation. By using the expression (\ref{npformexp}) in the equation (\ref{npdiffk}), similar to the one particle case, we find that the amplitudes corresponding to a particle $j$ being on the left and right sides of the impurity are related through the particle-impurity S-matrix (for ease of notation, from here on we suppress the spin indices unless needed)
 
 \bea \label{nprel}f^{...0j...}(z)=S^{j0}(z)\:f^{...j0...}(z).\eea
 
 Here again ``..." in the superscripts corresponds to any ordering of the rest of the particles. The particle-impurity S-matrix is given by

\begin{align} \label{smatk}&S^{j0}_{ab,\alpha\beta}(z)=\frac{ig(z_j)I^{10}_{ab,\alpha\beta}+P^{10}_{ab,\alpha\beta}}{ig(z_j)+1},\end{align}
 where the superscript denotes that it acts in the spin spaces of the particle `j' and the impurity `0'.  In the above expression $g(z_j)$ is related to the interaction strength $J(t-x_j)$ through the relation $g(z_j)= \frac{1}{2J(t-x_j)}\left(1-\frac{3}{4}(J(t-x_j))^2\right)$. Here we stress that this relation is non universal and depends on the specific regularization used. One obtains universal answers which agree with other regularization schemes in the limit of small coupling strengths $J(t)$ \cite{PasnooriRG}, where $g(z_j)\sim 1/J(t-x_j)$.  Note that we have ignored an overall phase in the S-matrix which is unimportant phase for our current discussion. In the above expression $I^{j0}_{ab,\alpha\beta}$ is the identity operator and $P^{j0}_{ab,\alpha\beta}$ is the permutation operator which acts in the spin spaces of the particle and the impurity and exchanges their spin \footnote{Explicit form of the permutation operator is
\be P^{ab}_{pq,rs}=\frac{1}{2}\left(I^{ab}_{pq,rs}+\sum_{\alpha=x,y,z}\sigma^{a,\alpha}_{pq}\sigma^{b,\alpha}_{rs}\right).
\ee
Here $\sigma^{a,\alpha}$ and $\sigma^{b,\alpha}$, $\alpha=x,y,z$ are the Pauli matrices acting in the spin spaces $a$ and $b$ respectively.}.

\begin{center}
\begin{figure}
\includegraphics[width=0.9\columnwidth]{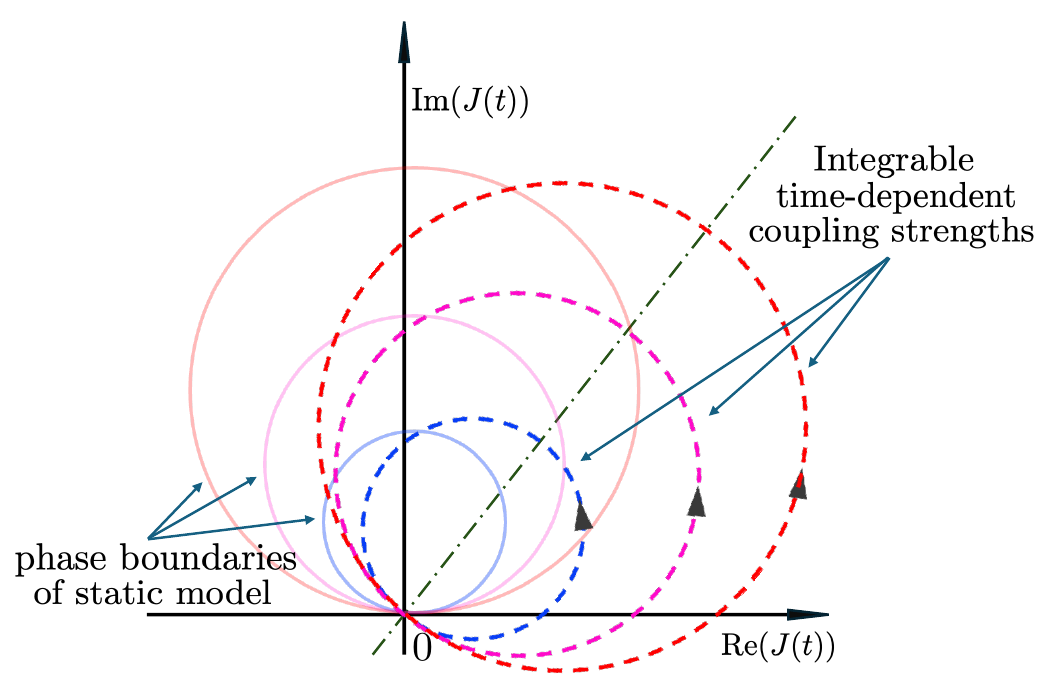}
\caption{Figure depicts the integrability preserving time-dependent strengths. The x and y-axes correspond to the real and the imaginary values of the time-dependent coupling strength respectively. The trajectories of the coupling strengths are perfect circles resting at the origin as shown (dashed circles) with the arrows representing the flow in the direction of increasing time. The points on a given circle that are on the right and left side of the origin are the past and future infinities respectively. All circles are centered on the slanted dashed line and the radius of each circle corresponding to an integrable time-dependent coupling strength is a dynamical invariant. When the rate of change of the imaginary value of the inverse of the interaction strength $1/J(t)$ is zero, the line formed by the centers of the time-dependent couplings is along the y-axis, which corresponds to the RG protocol. As the ratio of the rate of change of the imaginary and real value of the inverse of the coupling strength $1/J(t)$ is increased the slope of the line formed by the centers of the time dependent coupling strengths decreases, eventually merging with the x-axis where the rate of change of the real value of $1/J(t)$ is zero. The phase boundaries of the corresponding static model are represented by faint solid circles for comparison with the RG protocol. }
\label{fig:picture2}
\end{figure}
\end{center}

  As mentioned above, the consistency of the wave function required us to differentiate between the amplitudes which differ in the ordering of the particles with respect to each other. These pairs of amplitudes are not constrained by the Hamiltonian due to the relativistic dispersion \cite{PasnooriKondo}. To preserve integrability, one needs to choose a specific electron-electron S-matrix $S^{ij}(z_i,z_j)$
 that relates the amplitudes $f^{..ij..}(z_1,...,z_N)$ and $f^{..ji..}(z_1,...,z_N)$ which differ by the ordering of the two particles $i$ and $j$ with respect to each other 
 \begin{align}&
f^{...ji...}(z_1,...,z_N) =S^{ij}(z_i,z_j)f^{...ij...}(z_1,...,z_N),\label{eemateqsn}\end{align}
where
\begin{align}S^{ij}(z_i,z_j)=\frac{i\big(g(z_i)-g(z_j)\big)I^{ij}+P^{ij}}{ig(z_i)-ig(z_j)+1}.\label{smatee}
\end{align}
Here again ``..." in the superscripts corresponds to any ordering of the rest of the particles, which is the same in both amplitudes. The particle-impurity S-matrix (\ref{smatk}) and the particle-particle S-matrices (\ref{smatee}) satisfy the following Yang-Baxter equations

\begin{align}S^{j0}(z_j)S^{i0}(z_i)S^{ij}(z_i,z_j)=S^{ij}(z_i,z_j)S^{i0}(z_i)S^{j0}(z_j)\label{YB2},\end{align}

\begin{align}\nonumber
&S^{ij}(z_i,z_j)S^{ik}(z_i,z_k)S^{jk}(z_j,z_k)
\\&=S^{jk}(z_j,z_k)S^{ik}(z_i,z_k)S^{ij}(z_i,z_j).\label{YB3}
\end{align}
Using the relations (\ref{nprel}) and (\ref{eemateqsn}), one can relate any amplitude in the N-particle wavefunction (\ref{npform}) in terms of one amplitude of our choice. Without loss of generality, let us choose this amplitude to be $f^{N...10}_{\sigma_1,...,\sigma_N,\alpha}(x_1,...,x_N)$. This amplitude can be determined by applying periodic boundary conditions. Using the relation (\ref{pbcnp}), which is obtained by applying periodic boundary conditions, along with the relations (\ref{nprel}) and (\ref{eemateqsn}), one obtains the following equation
\begin{align} \nonumber &f^{N...j...10}_{\sigma_1...\sigma_N}(z_1,...,z_j-L,...,z_N)\\&=Z_j(z_1,...,z_N) \; f^{N...j...10}_{\sigma_1...\sigma_N}(z_1,...,z_j,...,z_N),\label{GqKZ}\end{align}

where

\begin{align}\nonumber 
    Z_j(z_1,...,z_N)=   S^{jj+1}(z_j,z_{j+1}+L)S^{jj+2}(z_j,z_{j+2}+L)\\...S^{jN}(z_j,z_N+L)S^{j0}(z_j)S^{j1}(z_j,z_1)...S^{jj-1}(z_j,z_{j-1})
.\label{transfermat}\end{align}

The operator $Z_{j}(z_1,...,z_N)$ transports the particle $j$ through the entire system once when acting on the amplitude $f^{N...10}_{\sigma_1...\sigma_N}(z_1,...,z_N)$. 

\paragraph{Consistency conditions and constraints on integrability}
In order for the system of equations (\ref{GqKZ}) to be consistent and hence have a solution, the operators (\ref{transfermat}) should satisfy the following consistency conditions \footnote{These conditions impose that transporting particle `j' around the system first and then particle `i' around the system is equivalent to transporting particle `i' around the system first and then particle `j' around the system.}

\begin{align}\nonumber
&Z_i(z_1,...,z_j-L,...,z_N) Z_j(z_1,...,z_N)\\&=
Z_j(z_1,...,z_i-L,...,z_N) Z_i(z_1,...,z_N).\label{commutations}\end{align}

These conditions impose constraints on the interaction strength such that 
\begin{align}
    g(z\pm L)=g(z)\pm \kappa,
\end{align}
where $\kappa$ is a constant. Working in the universal regime which corresponds to small values of the interaction strength $J(t)$, and using the relation between $g(t)$ and $J(t)$ (see below Eq (\ref{smatk})), we obtain the following integrable time-dependent interaction strength

\begin{align}
    \frac{1}{J(t)}=at+b, \;\; (a,b) \in \mathbb{C}.\label{intstrength}
\end{align}

\paragraph{RG protocol and beyond}
Let us now consider the case where $a$ is real. Differentiating (\ref{intstrength}) with respect to time we obtain 
\begin{align}
 \frac{d}{dt}\left(\frac{1}{J(t)}\right)= a.
\end{align}
for $a=1/\pi$, we see that this is exactly the RG equation (\ref{rg}) upon the identification of time with the logarithm of the cutoff $t=\log\Lambda$ \cite{Benhoare,PasnooriRG}. Thus we see that for $a$ being real, the integrability preserving time-dependent interaction strengths follow the RG trajectories, in agreement with the RG protocol. In the time-dependent model,  $Re(1/J(t))$ changes in time and corresponds to the running coupling strength in the static model, whereas $Im(1/J(t))$ is constant which is a dynamical invariant and corresponds to the RG invariant parameter in the static model discussed above.

We see from the expression for time-dependent strength (\ref{intstrength}) that $a$ can be complex. In this case, differentiating this equation with respect to `$t$' and identifying time with the logarithm of the cutoff does not yield the RG equation and in contrast to the RG protocol discussed above, here both the real and imaginary parts of the inverse of the interaction strength $1/J(t)$ change with time. Specifically, integrability imposes that both of these change linearly with time. Thus we find that in the non Hermitian Kondo model, the set of integrability preserving time-dependent strengths is larger than the set corresponding to the RG protocol. 

Similar to the RG flow as shown in figure (\ref{fig:picture1}), one can plot $Im(J)$ as a function of the $Re(J)$. Doing so, one obtains perfect circles like the limit cycles in the RG, but with the centers shifted away from the imaginary axis as shown in (\ref{fig:picture2}). One may naturally expect that when the correspondence between the integrable time-dependent strengths and the RG flow is applied to non-Hermitian systems, the limit cycles in the RG flow should correspond to periodic time-dependent strengths. It is interesting to note that even though the most general integrable time-dependent strengths do not follow the RG trajectories, they are still periodic.  In this general case, unlike the RG protocol although $Im(1/J)$ is not an invariant, it is interesting to note that the radius and the center of the circle traced by the time dependent coupling strengths (\ref{intstrength}) which are located at  $c=-a^{*}/(ab^{*}-a^{*}b)$, are indeed invariant in time. Hence we have identified new dynamical invariants in the time-dependent model, although their physical significance is not yet explored. We note that the time-dependent integrability allows one to change the couplings strengths at an arbitrary rate. It can be shown that in the case of slow driving one obtains the adiabatic regime where the system follows the instantaneous eigenstates \cite{pasnoorigrossneveu2}. In this regime, if we allow $Im(1/J(t))$ to change as described above, one can anticipate that the system allows one to remain in the instantaneous eigenstate but move across different phases described above \cite{pasnoori1}. In the most general case where the driving rate is arbitrary, one may expect to observe new dynamical phenomena, whose analysis goes beyond the scope of this present work.

In summary, we have considered the time-dependent non-Hermitian Kondo model. We have constructed an exact wavefunction using the generalized Bethe ansatz framework \cite{PasnooriKondo} and we have obtained the constraints imposed by integrability on the time-dependent strengths. In the time-dependent Hermitian models, these constraint equations on the time-dependent strengths take the same form as the RG equations of the corresponding static model upon the identification of time with the logarithm of the cutoff $t=\log\Lambda$ \cite{Benhoare,PasnooriRG}, which we refer to as the RG protocol. Here we have demonstrated that these constraint conditions in the non-Hermitian case are much more general: They include the RG protocol where time-dependent strengths follow the RG trajectories of the corresponding static non-Hermitian model. Additionally, they allow entirely new trajectories that do not match the static model's RG flow, thus demonstrating that the set of integrability preserving time-dependent coupling strengths is larger than the set corresponding to the RG protocol. We note that we have chosen to analyze the time-dependent non-Hermitian Kondo model since the corresponding static model has been well analyzed, but the construction provided in the current work is general and can be directly applied to other models with complex coupling strengths. Notable examples are the sine-Gordon model, chiral invariant Gross-Neveu model etc., where we find that the constraints imposed by integrability on the time dependent strengths are more general allowing time-dependent strengths to take trajectories that extend beyond those dictated by the renormalization group flows.

\section*{Acknowledgments}

We acknowledge enlightening discussions with J. D. Sau and S. Das Sarma. We acknowledge support from the Joint Quantum Institute and the Condensed Matter Theory Center, University of Maryland, College Park, where part of the work was done.

\bibliography{refpaper}

\end{document}